\documentclass[twocolumn,aps,amssymb,superscriptaddress,amsmath,prb]{revtex4-2}
\def\be{\begin{equation}}
\def\ee{\end{equation}}
\usepackage{graphicx}
\begin{document}
\title{THERMAL ENTANGLEMENT OF A GEOMETRICALLY FRUSTRATED SPIN-1 HEISENBERG DIAMOND CLUSTER}
\author{Azadeh Ghannadan} 
\affiliation{Department of Theoretical Physics and Astrophysics, Faculty of Science, P. J. \v{S}af\'arik University, Park Angelinum 9, 04001 Ko\v{s}ice, Slovakia}
\author{Katar\'ina Karl'ov\'a}
\affiliation{Department of Theoretical Physics and Astrophysics, Faculty of Science, P. J. \v{S}af\'arik University, Park Angelinum 9, 04001 Ko\v{s}ice, Slovakia}
\author{Jozef Stre\v{c}ka}
\affiliation{Department of Theoretical Physics and Astrophysics, Faculty of Science, P. J. \v{S}af\'arik University, Park Angelinum 9, 04001 Ko\v{s}ice, Slovakia}

\begin{abstract}
Thermal entanglement of a geometrically frustrated spin-1 Heisenberg diamond cluster is examined within the framework of the exact diagonalization method by computing the measure of entanglement negativity. The calculated exact analytical results are applied in order to obtain theoretical prediction of the robustness of bipartite thermal entanglement of the tetranuclear nickel complex [Ni$_{4}$($\mu$-CO$_{3}$)$_{2}$(aetpy)$_{8}$](ClO$_{4}$)$_{4}$ (aetpy = 2-aminoethyl-pyridine) against rising temperature and magnetic field.
\end{abstract} 

\maketitle
\section{INTRODUCTION}
Magnetic properties of a spin-1 Heisenberg diamond cluster were recently studied in Ref. \cite{KK_20}. It has been shown that the tetranuclear nickel complex [Ni$_{4}$($\mu$-CO$_{3}$)$_{2}$(aetpy)$_{8}$](ClO$_{4}$)$_{4}$ (aetpy = 2-aminoethyl-pyridine) \cite{Escuer}, which provides an experimental realization of the spin-1 Heisenberg diamond cluster, displays in a low-temperature magnetic curve one-half and three-quarter plateaus. In the present work the bipartite entanglement between two spin pairs either on a diagonal or a side of the spin-1 Heisenberg diamond cluster $\mathcal N_{12}$ and $\mathcal N_{13}$ will be quantified through the measure called the negativity \cite{Azadeh22}. The separability criterion of a state expresses that a state is entangled if and only if there is at least one negative eigenvalue of the partially transposed density matrix of the state, otherwise it is separable \cite{Peres}. Hence, the negativity is defined as the sum of absolute value of negative eigenvalues of the partially transposed density matrix \cite{VidWer}. 

\section{MODEL AND METHODS} 
The spin-1 Heisenberg diamond cluster is defined through the Hamiltonian:
\be
\hat{H}= J_{1}\,(\hat{S}_{1}\cdot\hat{S}_{2})+J_{2}(\hat{S}_{1}+
 \hat{S}_{2})\cdot(\hat{S}_{3}+\hat{S}_{4})- h\sum^{4}_{i=1} \hat{S}^{z}_{i},
 \label{Ham}
\ee
where $\hat{S}_{i}$ are the spin-1 operators, $J_{1}$ and $J_{2}$ are the coupling constants along the diagonal and side of the diamond cluster and $h$ is the magnetic field. The Hamiltonian (\ref{Ham}) was fully diagonalized in Ref. \cite{KK_20} and hence, the negativity can be calculated from a full set of eigenvalues and eigenvectors of the Hamiltonian (\ref{Ham}). The calculation details will be presented together with further results in our future work \cite{Azadeh22}. 
\begin{figure*}[t]
 \centering 
  \includegraphics[width=0.4\textwidth]{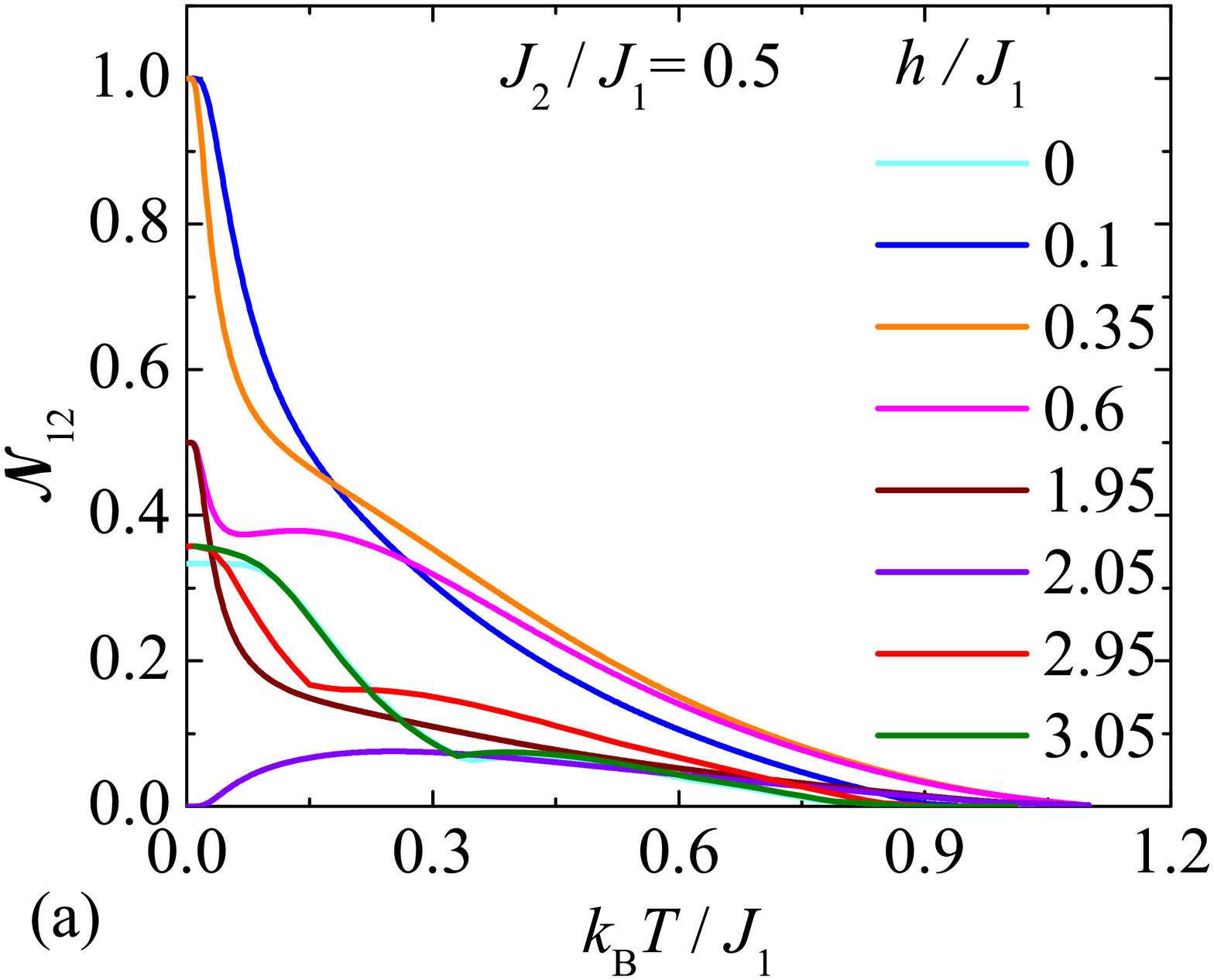}
  \vspace*{-0.3cm}
  \includegraphics[width=0.4\textwidth]{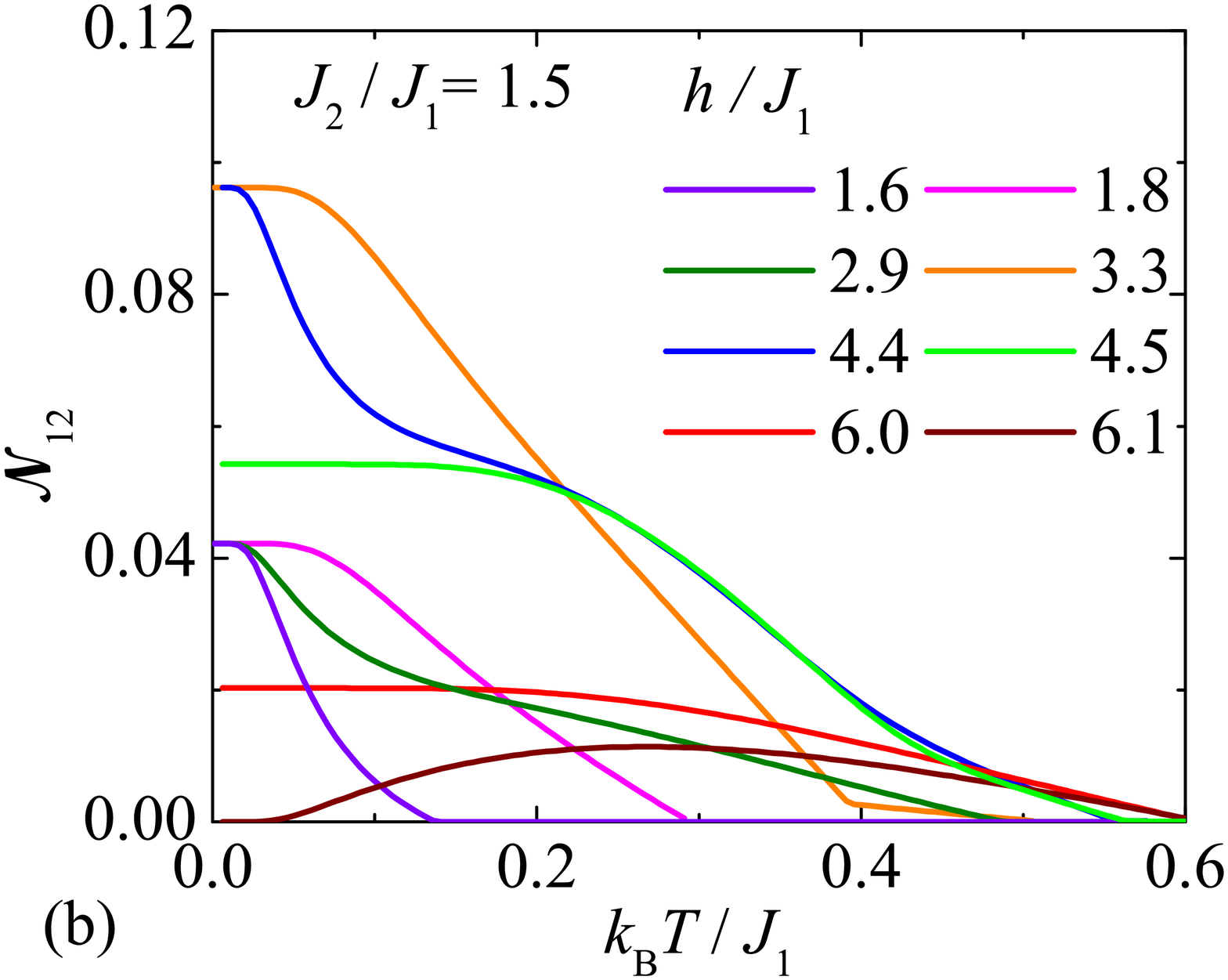}
  \hspace*{0.0cm}
  \includegraphics[width=0.4\textwidth]{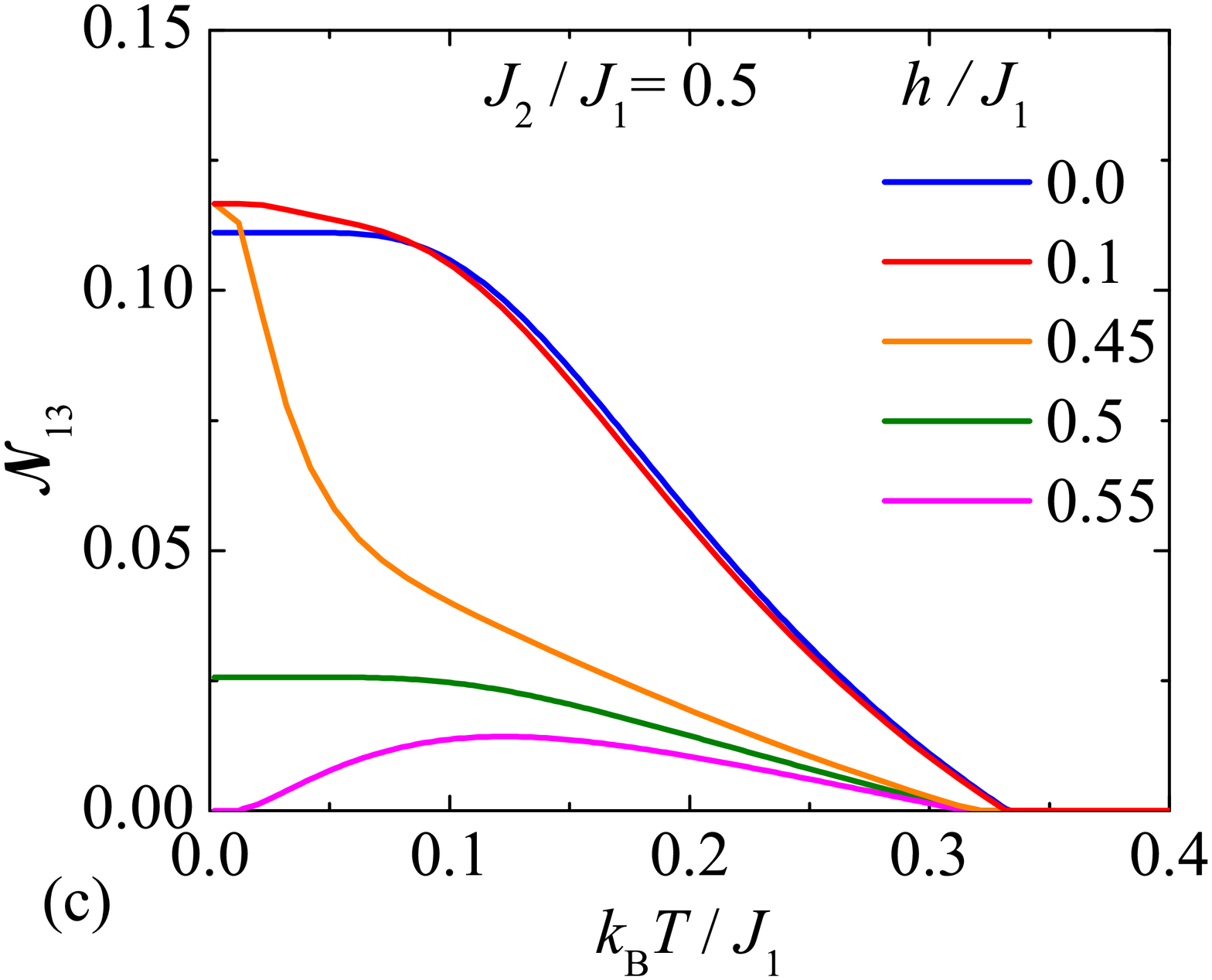}
  \vspace*{-0.3cm}
  \includegraphics[width=0.4\textwidth]{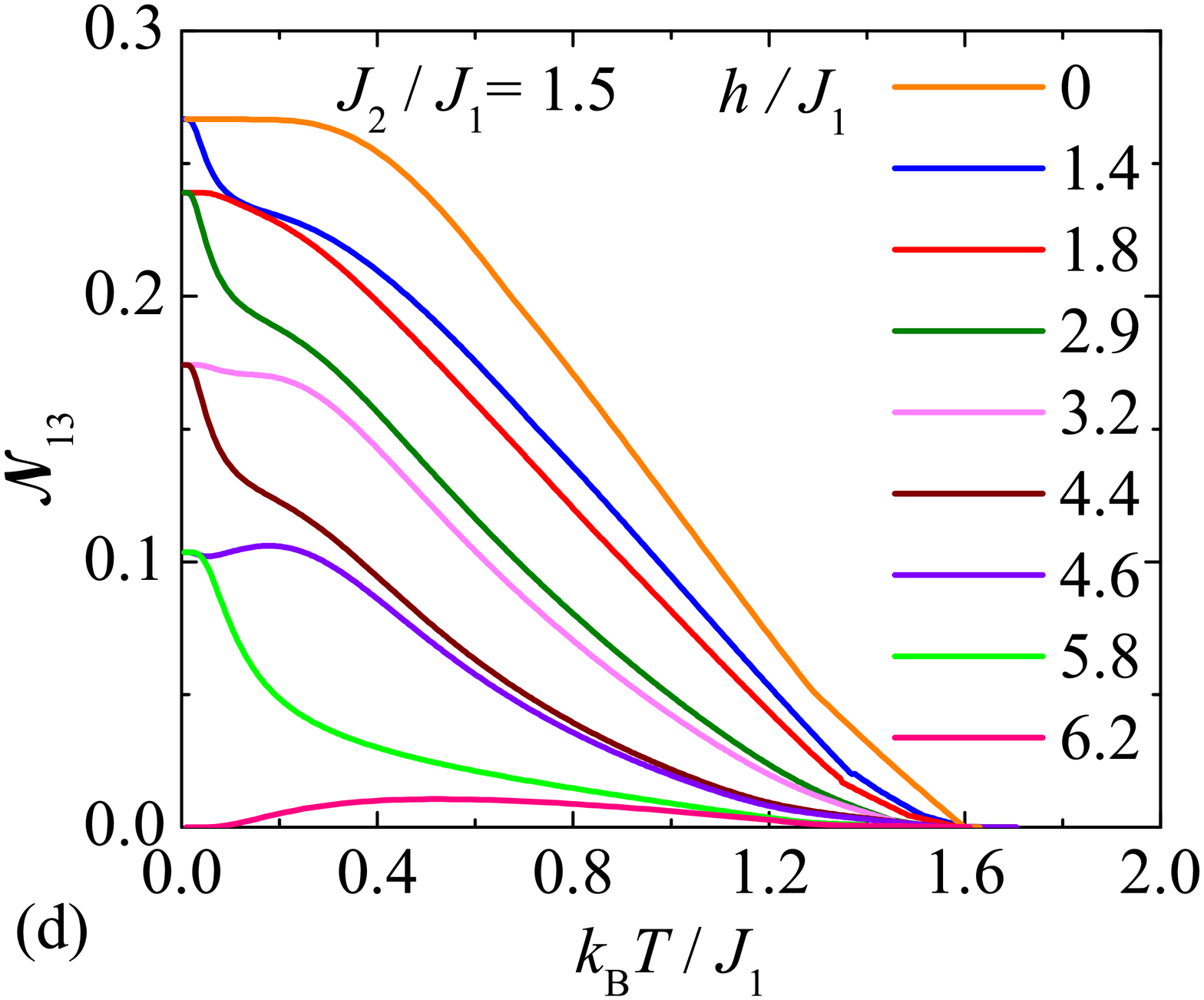}
  \hspace*{0.0cm}
	\caption{Temperature variations of the negativities $\mathcal N_{12}$ and $\mathcal N_{13}$ for the diagonal and side of the spin-1 Heisenberg diamond cluster for two different values of the interaction ratio: $J_{2}/J_{1}$=0.5 (a,c) and $J_{2} / J_{1}$=1.5 (b,d).}
	\label{fig:figure1}
\end{figure*} 
 To calculate the negativity one should first obtain the density operator and the corresponding density matrix of the given bipartite state. Then the density matrix is partially transposed with respect to one of its subsystems. Finally negativity is calculated as the sum of the absolute value of negative eigenvalues of the partially transposed density matrix according to the following formula \cite{VidWer}
 \be
 \mathcal N = \sum_{\lambda_{i}<0} |\lambda_{i}| = \sum_{i} \frac{|\lambda_{i}|-\lambda_{i}}{2}
 \ee

\section{RESULTS AND DISCUSSIONS} 
In this short paper we will discuss the thermal entanglement of the spin-1 Heisenberg diamond cluster. Thermal dependencies of the negativity $\mathcal N_{12}$ between a spin pair on the diagonal,  and the negativity $\mathcal N_{13}$ between a spin pair on the side are depicted in Fig. \ref{fig:figure1} for two interaction ratios $J_{2}/J_{1}=0.5$ and $J_{2}/J_{1}=1.5$. In both figures the curves starting from the same point in a zero-temperature limit $T$=0 imply the same ground state, note that the upper (lower) curve refers to the higher (lower) magnetic field. The curves with a single asymptotic value refer to a phase transition from one ground state to the other one. 
If the interaction ratio is $J_{2}/J_{1}=0.5$, the system goes through four different ground states upon increasing of the magnetic field, whereby $\mathcal N_{12}$=0 holds in one ground state and $\mathcal N_{13}$=0 in three of them. The results of nonzero negativities are exhibited in Fig. \ref{fig:figure1}(a) and (c). On the other hand the system goes through five ground states for $J_{2}/J_{1}=1.5$, whereby $\mathcal N_{12}$=0 holds in two ground states and $\mathcal N_{13}$=0 in one ground state. The results of nonzero negativities are shown in Fig. \ref{fig:figure1}(b) and (d).  

 It is obvious from Fig. \ref{fig:figure1} that negativity mostly decreases upon increasing of temperature until it terminates at threshold temperature. However, one may also a more striking re-entrance of the negativity when it is initially zero at low enough temperatures, then it shows a small temperature-induced rise until it repeatedly tends to zero at higher temperatures. A competition of high magnetic fields with slight thermal effects may thus cause a classical ground state to gain quantum properties. 
Another prominent feature which is observed in some thermal dependencies of negativity, is presence of the nonanalytical point known as a kink. The kinks occur when one or some negative eigenvalues of the partially transposed density matrix become zero at a given temperature. The kinks are also manifested in the corresponding magnetic-field dependencies of the negativity \cite{Azadeh22}. 

\begin{figure}[h]
  \includegraphics[width=0.4\textwidth]{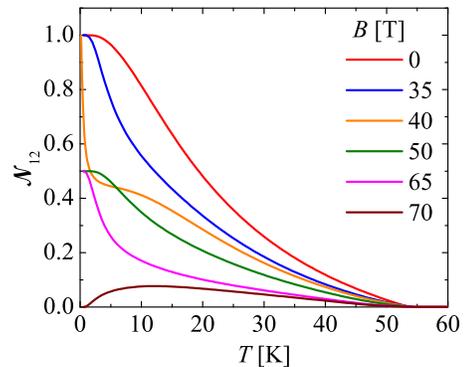}
  \vspace*{-0.5cm} 
  \caption{Temperature variations of the negativity $\mathcal N_{12}$ of the Ni$_{4}$ complex at a few different magnetic fields.}
	\label{fig:figure2}
\end{figure}

Now, let us shed light on thermal dependencies of the negativity of the tetranuclear nickel complex [Ni$_{4}$($\mu$-CO$_{3}$)$_{2}$(aetpy)$_{8}$](ClO$_{4}$)$_{4}$ (aetpy = 2-aminoethyl-pyridine) to be further abbreviated as the {Ni$_{4}$} complex, which affords a geometrically frustrated compound modeled by the Hamiltonian (\ref{Ham}). High-field data measured at low enough temperature $T$=1.3 showed presence of one-half and three-quarter plateaus in the magnetization curve of the {Ni$_{4}$} compound. Moreover, the values of the coupling constant between the nickel ions on the diagonal and side of Ni$_{4}$ complex are known $J_{1} /k_{B}$=41.4 K and $J_{2} /k_{B}$=9.2 K respectively. The coupling constant ratio $J_{2} / J_{1}$ =0.222 accordingly falls into a parameter region, where the system goes through three different ground states upon increasing of the magnetic field and entanglement exists just between the spin pairs on the diagonal. Fig. \ref{fig:figure2} displays temperature dependencies of negativity $\mathcal N_{12}$ of the Ni$_{4}$ complex, which is fully entangled $\mathcal N_{12}$=1 from $B$=0 $\rm {T}$ up to $B=40 \rm{T}$ due to formation of two singlet bonds within the diagonal spin pair. If the magnetic field ranges from 40 $\rm{T}$ to 65 $\rm{T}$ there exists another ground state, in which one singlet bond breaks and $\mathcal N_{12}$=0.5. All descending curves of negativity terminate in the threshold temperature about 55 $\rm {K}$. Beyond 65 $\rm {T}$ all four spins are directed to the magnetic field within the classical fully saturated ground state with $\mathcal N_{12}$=0. It is clear from Fig. \ref{fig:figure2} that the interesting re-entrance of the negativity occur for the magnetic fields slightly higher than 65 $\rm {T}$. 

\section{CONCLUSIONS}
In summery we have investigated in detail temperature dependencies of the negativity of the spin-1 Heisenberg diamond cluster for two different spin pairs on the diagonal and the side of the diamond cluster, which typically monotonically decrease upon rising temperature until they vanish at the threshold temperature. The striking re-entrance of the negativity was observed at high enough magnetic fields when relatively weak thermal entanglement can be induced above the classical fully polarized ground state by thermal fluctuations. Besides, the threshold temperature of the negativity of the Ni$_{4}$ complex was conjectured about 55 $\rm {K}$. 
\\
\begin{acknowledgments}
This work was financially supported by the grant Nos. APVV-20-0150 and VVGS-PF-2022-2101. 
\end{acknowledgments}

\end{document}